\documentclass[12pt]{article}
\usepackage{graphicx}
\usepackage{epstopdf}
\usepackage{amssymb,amsmath,amsfonts,palatino,amsthm}
\usepackage{amssymb}
\usepackage{epstopdf}
\newcommand{\beq}{\begin{equation}}
\newcommand{\eeq}{\end{equation}}

\newcommand{\f}{\begin{equation}}
\newcommand{\ff}{\end{equation}}

\begin{document}

%%%%%%%%%%%%%%%%%%%%%%%%%%%%%%%%%%%%%%%%%%%%%%%%
\title{Time and symmetry in models of economic markets}
\author{
Lee Smolin\thanks{lsmolin@perimeterinstitute.ca}
\\
\\
Perimeter Institute for Theoretical Physics,\\
31 Caroline Street North, Waterloo, Ontario N2J 2Y5, Canada}
\date{\today}
\maketitle

\begin{abstract}

These notes discuss several topics in neoclassical economics and alternatives, with an aim of  reviewing fundamental issues in modeling economic markets.  I start with a brief, non-rigorous summary of the basic Arrow-Debreu model of general equilibrium, as well as its extensions to include time and contingency.  I then argue that symmetries due to similarly endowed individuals and similar products are generically broken by the constraints of scarcity, leading to the existence of multiple equilibria.   

This is followed by an evaluation of the strengths and weaknesses of the model generally.  Several of the weaknesses are 
concerned with the treatments of time and contingency.  To address these we discuss a class of agent based models\cite{partecon}.  

Another set of issues has to do with the fundamental meaning of prices and the related question of what the observeables of
a non-equilibrium, dynamic model of an economic market should be.  We argue that these issues are addressed by formulating
economics in the language of a gauge theory, as proposed originally by Malaney and Weinstein\cite{MW}.  We review some of their work and 
provide a sketch of how
gauge invariance can be incorporated into the formulation of agent based models.

\end{abstract}
\newpage
\tableofcontents
\vfill

\section{Introduction }

Economics is a unique subject in that it is about nothing but human behavior, but it is also highly mathematical.  The economic theory that appears to be most widely in use by experts, called neoclassical economics or general equilibrium, is remarkably like physics, in that it is based on a few simple principles, which lead to  a rigorous mathematical formulation.  This mathematical formulation leads to some basic theorems, which are taken to be a rough or approximate description of a well functioning economy\cite{AD, Starr}.   A large body of results and models is built up around this theory, and there is a community of experts that express confidence in its basic correctness.  

At the very least,  the neoclassical theory of general equilibrium establishes that economics is  one of the mathematical sciences.  It sets a standard for clarity, rigor and generality that alternative approaches to economics must aim to live up to.  This theory is a model that mathematical scientists in other disciplines should be interested to study and understand, because there is a serious claim it has  succeeded in capturing something true in a simple mathematical structure, not about quarks or gravity waves, but about human beings. 

But if the evolution of the other mathematical sciences is any guide, one cannot expect that the first successful theory in a domain is the last word.  Precisely because of its success, we should expect that sooner or later any mathematical theory of real phenomena is replaced by a deeper and even more successful theory.  

There are reasons to think that the present is a perspicuous moment for seeking to improve economic theory.   There are prominent voices criticizing it, or its offshoots in theories of pricing and risk in financial markets, as having badly failed as a basis of policy advice in the last decade leading to the current economic crisis.  Indeed, the relationship between the actual theory of general equilibrium, which is claimed to be foundational for economics, and the advice given by professional economists is not completely clear.  Nonetheless it appears to be the case that there is a view among some economists that the model of an economy expressed by the theory of general equilibrium does teach lessons that have  implications for policy questions.  Typical among such views is the claim that markets do not require more than minimal regulation because they naturally evolve to a state of stable equilibrium, in which every participant is more satisfied and better off than they would be were the market constrained by external regulation\footnote{To anticipate one conclusion of the following, it does not appear to be the case that an objective reading of the mathematical theory of neoclassical economics provides an argument for laissez-faire economics over a mixed economy or for limiting regulation of financial markets.}.

There are also long standing critiques of the neoclassical notion of general equilibrium, some coming from academic economists, others coming from scientists in other disciplines.  Typical among these is the claim that the economic concept of equilibrium is at best incomplete and should be replaced by a notion of a steady state in a non-equilibrium, self-organized system\cite{SOC}.  This latter critique leads immediately to a confusion, because it is not true that the economic notion of equilibrium is at all related to the thermodynamic notion.  Nonetheless, once one gets past this confusion, there is still a cogent claim that non-equilibrium statistical mechanics may be a basis for a model of an economy.  This claim has to be proved on its own merits, and here it is fair to say two things.  First that there is an intuitive story which supports its plausibility, second, that nonetheless this idea has not led to a real theory that can rival the neoclassical notion of general equilibrium, either for its conceptual and mathematical elegance or for its wide applicability.  

Below, I will focus on two sets of issues which point to aspects of  the neoclassical model that could be improved upon\footnote{Needless to say, there are many interesting issues that are not discussed here, for example those having to do with limits on rationality of agents.}.  The first 
have to do with the treatments of time, contingency and dynamics\footnote{These issues turn out to be partly analogous to issues of time in  of physics and cosmology\cite{roberto,timeinphysics} }.  The Arrow-Debreu model can be extended to include a form of time and contingency but it appears only at the cost of entirely unrealistic assumptions, such as that each household can make a complete future projection of their earnings and consumption plans, for their whole lives, for all possible contingencies.  This seems adequate motivation to investigate models of markets where agents make decisions trade by trade, on the basis of the limited information available to them at each time.  
 
The second set of issues concerns the fundamental meaning of prices.  We will see that neoclassical economics answers this question in a way that makes it almost axiomatic that every good should have a price at every time.  Moreover there is only one price, for any good at any time, and all participants in the market know it and use it.  The price is determined by the condition of equilibrium, but there is no out of equilibrium dynamics which moves prices towards equilibrium. 

 But in real markets there must be such dynamics, and understanding how equilibrium arises-if it indeed does-from a more general non-equilibrium dynamics is a necessary step to answering many key questions about market behavior.  The question is then what is a price out of equilibrium?  Is there a single out of equilibrium price, defined absolutely, for all market participants, or is there a price for each transaction, determined by the interaction of those directly involved? 

Indeed, in  the agent based models of \cite{parteconI, parteconII},  prices arise dynamically from the trading activities between agents.  A price only exists after a trade has been completed and it is only known to the participants in that trade.  Whether and how a community of agents reach agreement on the price of a good is a dynamical question.  One way to express the difference is that in neoclassical economics the price of any good has an absolute meaning which is shared by all.  In the models of \cite{parteconI, parteconII} price is a relational dynamical quantity. (Thus there is an analogy to the role of space in Newtonian physics verses general relativity.)  To see the difference ask the question: {\it if a good has never been traded, never bought or sold, or has not been traded in a long time, does it have a price?}  Neoclassical economics says yes, the models of \cite{parteconI, parteconII} say no.  The difference is not just abstract, the issue of whether everything always has a price, even if it is not traded, is essential for disputes in accounting practices that have enormous practical import.  

Related to this are other fundamental issues about prices.  Is a price an absolute quantity, or is it a ratio?  In physics we have a basic rule which is that observables are always expressible as ratios of two quantities expressed in the same units, so  they are a pure number.  Applied to economics, this rule suggests that decisions should not be based on the units or currencies prices are expressed in.  Is there a way this can be enforced in the formal treatment?   For example, in a currency market, are the dynamics driven by pairwise ratios of goods or currencies, ie so many euros per dollar?  This seems unlikely to be meaningful as these are ratios of quantities in different units.  A better observable might constructed from  measuring the result of performing a cycle of trades which begin and end in the same currency.  Here one can take the ratio of two prices in the same currency.   However, there is a  problem, which is that all such quantities are supposed to vanish in equilibrium.  But perhaps this is what we want for the non-equilibrium theory.  Perhaps  these are quantities whose dynamics is responsible for the evolution from non-equilibrium to equilibrium.  

It turns out there is a natural approach these questions, which is that there should be in economics a notion of gauge invariance.  It expresses the fact that the real observeables with which dynamics is formulated are compound quantities, such as ratios.  Gauge invariance plays a fundamental role in phyiscs, because it addresses a similar set of questions concerned with uncovering the real observeables of the system.  The proposal that gauge invariance is fundamental for economics was made originally by Malaney and Weinstein\cite{MW}, and a brief summary of part of their work is given below in section 6.2\footnote{Other discussions of the role of
gauge theories in finance or markets are in \cite{other}.}. 

In fact, as we will see shortly, there are already gauge invariances built into the Arrow-Debreu model, although they are not usually labeled as such.  Our contention, following Malaney-Weinstein, is that this is the tip of an iceburg of a larger group of gauge symmetries that can be hypothesized to characterize a non-equilibrium model that may extend the Arrow-Debreu model.  

The aims of the following are modest, as they must be because the author is new to the subject.  There is a great danger in coming from outside an established field and believing one can contribute anything that is not trivial or nonsense.  Before doing so one should at least study the prevailing theory and the history of the field in some detail.  What follows is in part  a set of notes made while engaged in such a study.   At the same time, it is not impossible that this is a useful exercise, because there is a unity to the mathematical sciences.  This comes about by our use of the same mathematics and similar methodologies.  Indeed, there is an admirable  tradition of mathematicians and physicists contributing to the foundations of economic theory, such as Bak, von Neumann, Nash and Mandelbrot.  

Our own work in this area has been inspired by and carried out in the context of a collaboration, whose aim is to contruct ab initia agent based models of markets, in order to study basic questions about how markets work\cite{partecon,parteconI,parteconII}.  
Everything that follows is then to be considered tentative, a first stab at a very difficult set of problems.  This having been said, what is done in here is the following.  Section 2 is devoted to a heuristic account of the general equilibrium theory of Arrow and Debreu\footnote{This is based mainly on the classic paper of Arrow and Debreu \cite{AD} as well as a book of Starr\cite{Starr}.}.  Many details are left out, including some essential for stating the results rigorously.  The aim is to present only the very basics of the ideas, the basic mathematical formulation and the results. It is meant primarily for other physicists as an introduction. 

In section 3,  I give a heuristic argument as to why one would expect that real economies have a large number of states that satisfy the 
definition of equilibrium.  This is based on reasoning common in physics, having to do with spontaneously broken symmetries.  
Section 4 contains a critique of Arrow and Debreu, emphasizing both the strengths and weaknesses, the latter  paying special attention to the treatment of time.  It is likely that nothing here is original.  Section 5 then addresses the issue of the role of time, while section 6 reviews the proposal of \cite{MW} and explains why gauge invariance is fundamental for economics. In different ways these address the weaknesses of Arrow and Debreu's notion of economic equilibrium. Whether they will lead to theories that have any of the strengths of Arrow and Debreu's work remains to be seen.  

\subsection{What should we expect from models of economic markets?}

Before launching into the body a few notes are in order as to the aim of economic theory and models.  The reason is that while economics is like physics in some respects, in others it is very different.  Economics is by definition an approximate and highly idealized description of
a few aspects of the world.  In is entirely what physicists call an effective theory.  There is no fundamental description.  
An economic theory is intended to capture a few aspects of  very complex and dynamical emergent phenomena, which occur in large markets in human societies.  As such, there is no perfect model, the goal is to enunciate principles, which are general enough, but also powerful enough, to underlie the construction of a  variety of models adopted to specific circumstances and questions, and to be the basis of explanation and, to a limited degree, prediction.   Because no theory or model can perfectly capture the complexity of an economy, all principles, theories and models, must be considered provisional and phenomenological, and continually tested against data.  An attitude of modesty and open mindedness must govern our efforts in this difficult area.  

To the extent that there is a connection to other areas of science it comes about because the economy is a subset of human activity, which forms a subset of processes in the biosphere.  The biosphere is a far from equilibrium system, driven to an approximate, slowly evolving, steady state by the flow of energy through it.  It is characterized by cycles of material and energy whose rates of flow are governed by feedback processes acting on diverse scales.   

The economy of the world is a subset of that system of cycles and flows which is interesting to us because it is subject to peculiar kinds of decision making, about which we keep records.  The economy is, in fact, defined by the records we keep of it, it is not a naturally occurring or specifiable subsystem apart from our interest in the records that are kept of it.  

Even if we restrict attention to all the financial and economic records in the world economy we have a fantastically complex system.  So to make progress we make models, which are inevitably gross simplifications. 

Furthermore, an economy is subject to external events and conditions, which being human events are not subject to simple modeling or reliable predictions.  This point has been made different ways by several authors.   Taleb\cite{blackswans} has emphasized that unpredictable events have major influences on directions of markets and societies.   Kauffman has emphasized that economic growth and social change are driven by innovations which are for all practical purposes, if not in principle, unpredictable\cite{stu} .  Mangabeira Unger  has described how social change at all levels is driven by the creation of new modes of organization, into an ever increasing space of possibilities\cite{roberto}.  So there is a strong limitation in what we can expect from a formal economic theory.  But we still can investigate the possibility that there are some principles that lead to models, which can make some useful and testable predictions-at least for the behavior of markets between their disruptions by unpredictable events and innovations.  To the extent that these are robust and independent of the fine-grained details of how human actors make decisions our models will be reliable.   We are then interested in the simplifications that may happen in limits of large numbers, such as a large number of agents, or of goods. In these limits there may be universality classes that only depend on a few parameters that characterize the behaviors of the human actors that comprise the economy.  

Thus, we are inclined to side with the neoclassical economic theorists against critics who suggest that because human behavior is involved there are no useful regularities that can be captured by mathematical models of a market.   Even as we are mindful that there can be no perfect model of an economy, we take the Arrow-Debreu model as a first example of a useful model, and we take its shortcomings as starting points to attempt to improve it rather than as indications that there cannot be a useful general theory of economies and markets.

\section{The basics of neoclassical economic theory}

I give a brief and non-rigorous summary of the basics of neoclassical ecnomic theory as formulated by Arrow and Debreu and others.
We start with a pass at the basic ideas in words, then we will see how they are formalized.  

\subsection{The basic idea}

The basic subject of classical and neoclassical economics is how societies solve problems of the allocation of scarce resources.  The basic
questions to be answered by an economic society are\cite{samuelson}, {\it What gets made?}, {\it By whom and how does it get made?} and {\it Who gets what is made?}, all subject to 
a lot of constraints, having to do with the finiteness of available resources, including natural resources, energy and human
beings' work.   The basic idea of
a market economy is that most of these decisions are made by individual companies and individuals and families, each of whom
is trying independently to extremize some measure of their happiness.  In the case of companies the measure of happiness is usually taken to be profits.  In the case of {\it households} (the technical term for individuals and families) they are thought to be trying to do as
well as they can on a single measure of their happiness, which is usually called {\it utility}.  That is,  given a choice between any two collections of goods or services they might purchase with their available funds, they are assumed to have a preference.  

One then wants to solve the problem of whether it is possible for all these different companies and households to simultaneously 
extremize the measures of their happiness, subject to the overall constraints on resources, plus constraints on what technologies are available to companies and what funds are available to people.  Allocations of the resources and goods that satisfy this are called {\it efficient}.  (I give more precise definitions below.)

This is formally a problem of simultaneously extremizing a very large number of functions of a large number of variables, 
subject to an also large and non-linear set of constraints.  It is almost by definition a difficult problem about a complex system. 

There is another desirable property an economy might have which is that every good which is made is sold and every service
or skill offered is employed.  Furthermore there should not be any unsatisfied demand for goods,  services or labor. When this is the case it is said that {it all markets clear.}  A market where this happens is called
{\it in equilibrium.}   

The first observation of neoclassical economics is that there is an elegant solution to both problems of efficient and equilibrium economies,  which is to have a currency which can be used to buy and sell everything. Each good and service then has a price, which is not fixed but is allowed to vary as the companies and individuals bargain with each other.  The big idea is that there is a simple mechanism, which is the law of supply and demand, which leads to a solution of both problems.  This is that prices go up when demand exceeds supply and go down in the reverse case.  The big results, which I will state a bit more formally below are that 1) under reasonable conditions there are always choices of prices so that the market is in equilibrium and 2) the equilibria states are also efficient, in the sense that no one can be made happier without decreasing someone else's happiness.   

\begin{figure}[ht]
\centering \includegraphics[width=5cm]{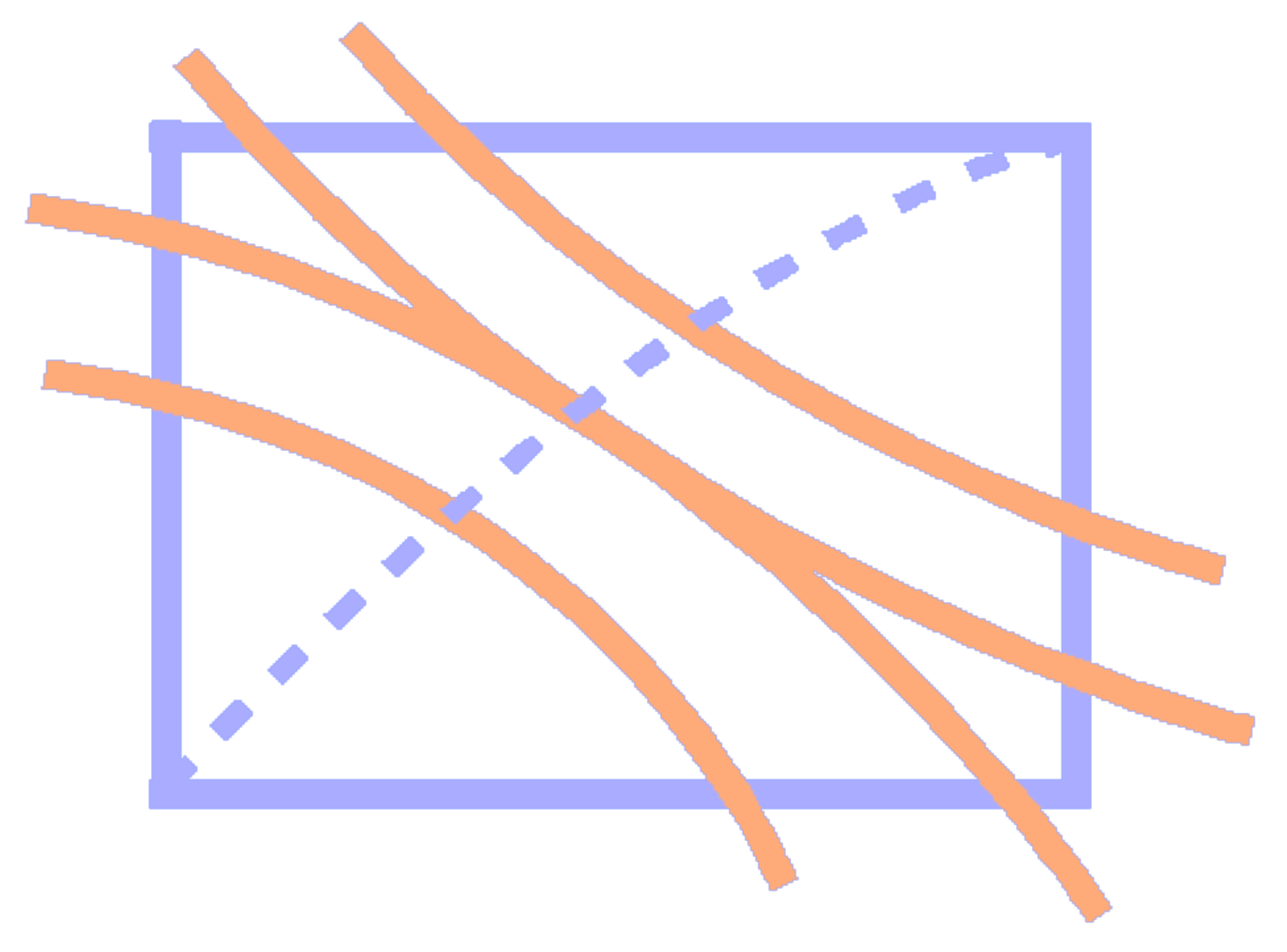}
\caption{The Edgeworth box.  The two axes are the amounts of two goods allocated between two people, subject each to the constraint
of a fixed total number of goods.  The curves are curves of equal happiness (or utility).  The set of curves convex to the origin are the level curves of the utility function of one person, the set convex to the corner opposite the origin are the utility function of the other person, which are valued in axes opposite those of the first person.  The point where they intersect and have equal tangent vectors is the equilibrium point, where no one can increase their utility without decreasing the utility of the other.  We see that convexity is necessary for equilibrium to exist, this expresses the principle of {\it diminishing returns,} ie two boats brings less than twice the happiness of one boat\cite{figcredit}.
\label{fig1}}
\end{figure}

To see why these are plausible results, one can consider a simple model with two people trading two goods.  This is something like the harmonic oscillator of economics and is called the Edgeworth box\cite{Starr}.  It is  illustrated in Figure (\ref{fig1}).  The total numbers of the two goods are fixed, the problem is how to allocate them between the two friends so each is as happy as possible.  Each has a function that determines how happy they will be as a function of how many of each good they have.  All allocations satisfying the constraints are points in the box.  It is not hard to see that there are points which extremize both people's happiness.
%..........................................................................
%..........................................................................

While this is instructive, it is highly non-trivial that similar results hold for markets with huge numbers of participants, many goods and many non-linear constraints.  The fact that they do under a general set of assumptions is the great success of neoclassical economics.  

Before going into the details it is important to raise and address an objection that is often given to the possibility of any successful mathematical economic theory.  This is that we are dealing with human beings whose behavior is changeable and indeed capricious, and human psychology rules how we behave in markets.  The result is that there are phenomena like bubbles and panics which may not be easily captured in mathematics.  

There of course is a lot of truth to this.  Nonetheless there is a cogent claim which stands as a first answer.  Whatever else is true there is a universal about human beings in markets, which is that they seek to maximize their happiness.  This, plus a few other such general assumptions, gets us to the claim that all the human beings together acting to extremize their happiness, results in a collective state, called equilibrium.  This state has some general characteristics, which can be formulated and studied without having to get much into what makes real human beings happy.  Even if the state of equilibrium is disturbed by panics, the claim is that this and other phenomena are transient, while what governs in the long term is the collective effect of the desire of every person to be as happy as their circumstances allow.  

This is a powerful argument, but let us point out that it rests on some assumptions about the state of equilibrium.  First, that it is unique, and second that it can be reached by a real market in a time scale short compared to that over which people's ideas about what makes them happy changes.  We will see below what is known about these questions. 

But now we are getting ahead of ourselves, lets now begin the formal development of these ideas.

\subsection{Basic kinematics of an economy in Arrow-Debreu}

We now see how these ideas have been formalized by neoclassical economists\cite{AD,Starr}.

\subsubsection*{The space of goods}

We start with the space of goods or products. This is an $N$ dimensional vector space, $\cal P$ which is defined to be ${\cal R}^N$. 
There is a prefered orthonormal basis, each element of which corresponds to a specific good or service or job, ie something
that could be bought or sold. These will be called, generally, goods.  
That is, the coordinates of a vector, $X^a \in {\cal P} $ corresponds to a quantity of goods of type $a$.  So such a vector can be
thought of as an inventory giving a list of goods that may be held.  

\subsubsection*{The treatment of time}

If it is relevant to model the economy at more than one time, then the same good, at different times, is considered a distinct
good.  That is, then the space of goods is taken to be ${\cal P}^T$ where $T$ is the number of distinct moments of time to be
modeled.  This means that each index is a pair $(a,t)$, where $a$ labels distinct goods and $t$ labels times.  

\subsubsection*{The price covector}

We assume the existence of a single currency.  Then there is a price co-vector, $\vec{p} = p_a = \{ p_1, p_2,...\} \in {\cal P}^*$ 
corresponding to prices of the $N$ goods, so $p_2$ is the price of the second good.  Each price is assumed 
to be non-negative, $p_a \geq 0$.   
The main problem we will be discussing are
conditions under which a preferred set of prices are determined.

Given an inventory, labeled by $X^a$ and a price covector labeled by $p_a$, the value of the inventory is
the contraction\footnote{{\bf Mathematical note}:  there is no metric on $\cal P$, so all meaningful quantities are 
contractions of production or consumption plans, which are vectors, with prices, which are covectors.  The natural metric on
$R^n$ appears to play no role.  This means that the relevent geometry is projective rather than metric, and leads
to the gauge symmetries to be discussed below.},  ${\cal V} = p_a X^a$.

We will see that the dynamics of the model are always homogeneous of degree one in the prices $p_a$.  This means
that there is a symmetry in which all prices are scaled by the same amount,  this corresponds to the independence of 
the dynamics from the units in which prices are valued.  This is the first hint of the role that gauge symmetry plays in economics, a point
we discuss in more detail below.  
This point is  obscured in the formulation of the Arrow-Debreu model because this symmetry is immediately eliminated (by what 
physicists would call a gauge fixing) by normalizing the  prices, so that 
\f
\sum_a p_a =1 ,
\label{normalize}
\ff  
This defines a simplex ${\cal S} \subset {\cal P}$ which is the space of prices.

\subsubsection*{Firms}

The economy is assumed to have $F$ firms or businesses.  Each one's contribution to the economy is described by a
{\it production process} which is a vector $Y_A^a$ where $A=1,...,F$ labels the different firms and $a$ labels the different goods.
The elements of the vector, if negative, denote an input to the process, that is if $Y_1^{17}= -4 $ then four units of good $17$
are inputs to the process of firm $A=1$.  When an element of the process vector is positive that is an output of the process, 
i.e.  if $Y_3^7= 6 $ then six units of good $7$
are outputs of the process of firm $A=3$. 

For each firms there is a prespecified set of possible, or accessible production processes, given by a 
compact, convex set ${\cal Y}_A \in {\cal P}$.   The meaning of the convexity requirement is that if
$Y_A^a \in {\cal Y}_A$ then so is $\lambda Y_A^a \in {\cal Y}_A$ for $0 < \lambda \leq 1$.  That is, if a firm 
can produce $10 $ cars it could choose to produce three.

For each set of prices, $p_a$ and process $y_A^a \in {\cal Y}_A$ the profit of that process at those prices
is given by $p (y)_A = p_a y^a_A$ which is a function on ${\cal Y}_A$.  

A firm can be assumed to act always to maximize its profits, given the opportunities and prices, so let
$y_{A *}^a \in {\cal Y}_A$ be the accessible process which maximizes the profit, as a function of the prices.
This is a mapping  from the space of prices, $\cal S$ to $\cal P$.  This function is called the 
{\it supply function},  $S^a_A (p)$ of the firm $A$. 

\subsubsection*{The household or consumer}

There are $H$ households, labeled by $\alpha = 1,..., H$.  
A consumer or household is characterized by a {\it consumption plan}, $X_\alpha^a$ which for each household, labeled, $\alpha$,
is a vector in the positive definite quadrant of $\cal P$, which we will call ${\cal P}_+$.  Its components, positive numbers,
represent a plan to consume each good. That is $X_{13}^5=12$ represents the intention of household $13$ to 
consume $12$ units of good number $5$.  We note that the cost of this as a function of the prices is
$p_a X_A^\alpha$.   

A household is also characterized by an {\it endowment}, $r_\beta^a \in {\cal P}_+$.  These are goods and services it can sell, including
the labor of its members.  The assumption is made that a household will sell all of its endowment, earning an income of
\f
i_\beta =p_a r_\beta^a.
\ff  
Note that this means that if  a member of a household wants to work less than the legal maximum, if there is one,
this is considered to be purchased back in their consumption plan, and is then called {\it "leisure"}.   

A household may also hold shares in a company, which are denoted by $\alpha_A^\alpha$.  This means that the income
of household $\beta$ is augmented by 
\f
i_{shares}^\beta = \sum_A \alpha_A^\beta p_a y^a_{A *}
\ff
Hence, the total income of the household is 
\f
I_{total}^\beta = p_a r_\beta^a + \sum_A \alpha_A^\beta p_a y^a_{A *}
\ff

The choice of a consumption plan is determined by a preference ordering, unique to each household, of the possible
consumption plans.   That is, each household has a partial order, denoted, $\geq_\beta$ on ${\cal P}_+$.   

Equivalently
each household has a utility function $U_\beta $ on ${\cal P}_+$, such that 
$U_\beta (X_\beta^1  ) \geq U_\beta (X_\beta^2  ) $ iff $ X_\beta^1\geq_\beta X_\beta^2$.   The utility functions
are then defined up to monotonic rescaling, so that utility functions of different households are not comparable, nor
is the numerical value or absolute differences between different utilities meaningful.  

Given prices $p_a$ there is a domain $\tilde{\cal P}(p)_+^\beta$  of consumption plans which the household $\beta$ can afford,
given their endowment and share income.
These are those such that 
\f
p_a X_\beta^a \leq I_{total}^\beta 
\ff 
The household will be assumed to maximize their preference of consumption plans within this set.  This means that 
they pick the $X^a_{\beta *} \in \tilde{\cal P}(p)_+^\beta$ which maximizes their preference ordering, or equivalently
maximizes their utility function.  This gives, for each household,  another map from the simplex of prices, $\cal S$ to 
$\tilde{\cal P}_+$, which is called the {\it demand function}.  Labeled $ D(p)_\beta^a$ it is equal to the consumption plan
$X^a_{\beta *} $ that maximizes their preferences or utility at a given set of prices.

\subsubsection*{The excess demand function}

We now give mathematical expression to the notion that supply matches demand.  This is expressed by means of
an excess demand function, which a map from prices to elements of $\cal P$ defined by
\f
Z^a (p) = \sum_\beta D(p)^a_\beta  - \sum_A S(p)_A^a -\sum_\beta r_\beta^a
\ff

For a given good, $a$, this is the amount that the overall demand from all households and inputs to processes exceeds
the supply given by endowments and outputs of processes, at a given price, $p$.  

\subsection{The notion of economic equilibrium}

To summarize, the kinematical state of an economy is then given by the choice of a production process, $y_A^a$ for each
firm, a consumption plan $X_\beta^a$ and endowment $r_\beta^a$ for each household, and an overall set of prices.  
The dynamics of each economy are determined by the choice of goods, the possible production plans of each firm and the
preference functions of each household.  Given those, we want to explore proposals for dynamical principles which will 
lead to the choice of kinematical state.  

Given the prices we can maximize profits for each firm, at each set of prices and within the set of allowed processes, to find that firms supply function, $S(p)^a_A$.  Similarly, we maximize each household's preferences within the set of plans, they can afford 
(which depends on prices as well as production plans-through their stocks) to find their demand function, $D(p)^a_\beta$.  

The idea of equilibrium in this context is that there will be prices where, when every firm has extremized their profit within constraints, and every household their preferences,  the supply precisely balances demand for each good\footnote{A very good critical review of the notion of equilibrium in 
economics and finance is \cite{FG}.}.   This means that the 
excess demand function,  $Z^a (p) $, vanishes.  In this heuristic
treatment we will ignore complications and technicalities, (such as what happens if the price of some goods vanish.) 

Thus we say that $p_a^*$ is an equilibrium price vector whenever 
\f
Z^a (p^*) =0
\ff

That is, these are sets of prices for which, within the constraints that each firm and each household are subject to,
each firm has maximized profit and each household has chosen their most preferred consumption plan.  

Note that the extremizations leading to the definition of the supply and demand functions for each participant are done separately
for each price.  This then assumes that no decision or action of a firm or household can influence the price.  The assumption is that all 
possible decisions of individual participants involve infinitesimal proportions of the goods, and so cannot change the prices.  Nor 
can there be, on this assumptions, any associations, monopolies, unions or guilds that give a coalition of participants the power to
change prices by their decisions.  This is called the assumption of {\it perfect competition.} 

Away from equilibrium $Z^a $ does not vanish because supply and demand are not balanced.  But a weaker property does hold out of equilibrium, which is Walras's law.  
\f
p_a Z^a (p) =0
\ff
This asserts that given any prices, the total cost of the unsatisfied demand is equal to the total cost of the unsold supply. The justification of
this law is given in \cite{Starr}, intuitively it expresses the fact that under the assumptions given, in order to maximize their utility at all prices,  all the endowment and income of households is spent. So at any price, the money available to consumers to maximize their utilities is determined by the profits made by the production plans that maximize profits. 

There are several things one wants to know about these equilibria.

\subsubsection*{Existence}

There are very general results that show the existence of at least one set of equilibrium prices for each such formal economy.  
This is the famous Arrow-Debreu-Mackenzie theorem.  

The idea of the proof is remarkably simple.  We can define a map from the simplex of prices to itself, as follows.
Given a price, $p_a \in {\cal S}$ compute 
\f
T(p)_a = \frac{p_a + \mbox{max} [0, Z_a (p)] }{1 + \sum_a\mbox{max} [0, Z_a (p)] }
\ff
This gives a map from ${\cal S}$ to itself.  It does the intuitive thing of raising prices of goods where demand exceeds supply
and lowers the price of goods where supply exceeds demand.  

The simplex of prices is compact and the map can be proven to be smooth. By the Brower fixed point theorem the map will
have at least one fixed point.  This is a $p^*_a$ such that $T(p^*)= p^*$.  This then implies that $Z^a(p^*)=0$.

\subsubsection*{Optimality}

The equilibria so defined can be proved to be optimal in the following sense.  A consumption plan for each household
is said to be Pareto efficient if, given a fixed set of prices and the corresponding allocation of goods that extermize
all the utility functions at those prices,  any re-allocation of the goods will lower at least one households utility. 
That is,  consumption plans are not Pareto efficient  when ther are reallocations of goods, which are allowed under the
constraints that define the economy, such that everyone's preference order or utility function is left the same or increases.

The connection to the notion of equilibria is the captured by a theorem called the {\it first fundamental theorem of
welfare economics.}  It says that if $p^*$ is an equilibrium price vector, then the associated supply and demand
functions define a set of consumption plans for each household which are Pareto efficient.  

There is also a result which is in the direction of a converse, called the {\it second fundamental theorem of
welfare economics.}  Roughly speaking it says that given any allocation of goods amongst households which is Pareto efficient,
there is an allocation of endowments among the households such that there are prices such that that allocation is also 
in equilibrium.  

\subsubsection*{Stability}

Little appears to be known about the stability of generic equilibria.   It must be the case that many of the large number of 
equilibria which exist are unstable, but one would like to know much more.  

For example, here is a line of thought suggested by Roumen Borissov.  It may be wrong, but if so it would be good to have
good evidence that it is wrong.  

Given some basic insights from complex systems, one
might even imagine that equilibrium points are generically unstable.  This is because they may correspond to points on boundaries in the state space 
between regions of order and chaos.  In the ordered regime fluctuations are small and gaussian, in the chaotic regime dynamics is chaotic and there are no fluctuations around stable points.  Points on the boundary have power law fluctuations which means that large excursions are common.  In such a state measurements of volatility are not meaningful because moments of the distribution are not defined by finite sets of measurements.  

If this is the case then we may not want a national or world economy to be at an equilibrium point.  Instead, there will be nearby states, which are not quite as efficient or balanced between supply and demand, where the fluctuations are smaller and gaussian, so that measures of volatility are meaningful and reliable.  Given this, we might prefer to design an economy so it is at a stable point, which means we sacrifice  efficiency and complete market clearing for a lot of stability and reliability.  

\subsubsection*{Uniqueness}

There is considerable evidence that the equilibria sets of prices for generic economies are highly non-unique.  I 
will give an argument to this effect in section 3.  However, there is a much more powerful result, which is the Sonnenschein-Mantel-Debreu Theorem\cite{SMD}.  This basically says that for any finite set of points of $\cal S$, there is an economy, defined as above, such that these are the
equilibria price vectors of that economy.  That is, any finite set of prices, no matter how many, can be the equiilibria prices for some
economy.  This is sometimes called the "anything goes theorem."  At the very least it suggests that generically equilibria of
economies are highly non-unique\footnote{It is interesting to speculate that the market may exist in a state which flows through many equilibria, as in a spin-glass\cite{spinglass}.}.  

\subsubsection*{Finding equilibria}

One would like to know how likely it is that an economy starting off at an arbitrary state, reaches equilibrium after a relatively short time.
It is possible to imagine that while the equilibrium prices exist, the processes at work in a real economy do not lead prices to converge,
quickly, if ever, on the theoretical equilibrium prices.  This is called the problem of tatonnement and it is discussed in \cite{FG}.

\subsubsection*{Invariances}

There are two invariances in the Arrow-Debreu model.  These are gauge invariances, in the sense that different mathematical representations correspond to the same economic model.  The model is invariant under the following
mathematical operations.  

\begin{itemize}

\item{} Rescaling the prices, so 
\f
p_a \rightarrow \Lambda p_p,
\label{scalingp}
\ff
where $\Lambda >0 $.  This is a global gauge invariance, and it is gauge fixed by imposing the condition (\ref{normalize}). 

One may criticize the assumption that (\ref{scalingp}) is a gauge invariance, noting that $\sum_a p_a$ does in fact change in time in real economies.  This is, however, not quite germane, because, as a good at different times is a different good, the sum over all prices, includes a sum over all times, hence what is normalized by the gauge fixing condition (\ref{normalize}) is the sum over all prices at all times.  What one wants to posit is that there may be a symmetry in which all prices at a single time may be scaled, in which case this is likely a global symmetry.    It is then interesting to ask whether there is a conserved current associated
with this symmetry\footnote{It has been hypothesized that there is a Goldstone mode related to the spontaneous
breaking of this invariance, which is inflation\cite{per-inflation}.}.  

\item{} Rescaling the utility functions
\f
U_\alpha \rightarrow \lambda_\alpha U_\alpha
\label{scalingU}
\ff
for $\lambda_\alpha > 0$.  
This can be considered to be a local gauge invariance because the utility function of each household can be scaled separately to reflect
the idea that the relative amounts of utility of different households are not comparable.  

\end{itemize}

Thus, the gauge group of the Arrow-Debreu model is $R^{H+1}_+$.  
We will argue below that this may be a subgroup of a larger gauge invariances of non-equilibrium models.

\section{Symmetry and multiplication of equilibria}

It is easy to use arguments based on symmetry to show that in many specific models there will be multiple equilibria.  All one has to
do is arrange it so there are symmetries amongst households or firms which dynamically must be broken because there are
not enough goods, or jobs, for everyone to get the same choice.

For example, suppose there are $n$ households who want a soft drink at lunch today in Alice's restaraunt, but are indifferent as to whether it is a Pepsi or a Coke, or even how much is one and how much is the other.  There is then a symmetry in their utility functions, which has the geometry of a line segment to the $n$'th power.  But suppose Alice has only $l < n$ cans of Coke available and $n-l$ cans of Pepsi.  In the attainable Pareto efficient states  $l$ of the $n$ have Coke and the rest have Pepsi.  There are $n \choose l $ such states.  If any of them is an equilibrium, so are all of them.  

Or suppose there are $n $ customers who want a cell phone with a certain list of attributes and $p$ companies which make them.
As they all have the same capabilities, the utility functions of the $n$ customers are indifferent to which of the $p$ cell phones
they will purchase.  But now suppose that there are exactly enough, so that each company makes $n_i$ phones such that
$\sum_i n_i =n$.  Then there are a large number of equally Pareto effecient states in which these phones are in the end
purchased by the $n$ customers.  Again, if any of them is an equilibrium, all of them will be.

Let us take a more serious example.  Suppose there are $n$ very good art students who all want a place in a New York gallery, and 
suppose there are $p << n $ actual places.  The market allocation mechanism will somehow choose $p$ of the $n$ young
artists and place them with a gallery, after which they will have (this is a model!) big careers with lots of income which will enable
them to consume lots of goods.  But the majority of the $n$ artists will have careers as art teachers, where they will have
modest incomes.  Each of the $n$ students has initial allocations which include their availability to be employed as teachers and the paintings they will make in their careers, which we will assume will only sell if they are in the minority chosen by a gallery.

Now tastes differ, but I will assume for the sake of the model, that all $n$ are considered good artists, so that the dealers would 
each be happy to have as many as they can take.  But they have in total only $p$ places.  
So there is a symmetry initially which is again $n \choose p $. This symmetry will necessarily get broken by the dynamics of whatever
market allocation mechanism functions to find Pareto efficient states.  All states in which any $p$ of the $n$ artists are chosen
by galleries can be Pareto efficient, that is the galleries will maximize their profits in each case and the artists will all, whether they
are stars or teachers, maximize their utilities subject to the constraints on their income.  Hence, if any of these are equilibria,
all of them are.  

I could multiply these kinds of examples, but the point is clear.  The situation in which the actual allocation of goods or jobs
in society is forced by scarcity to break a symmetry in the production and utility functions is absolutely typical.  One may try to
argue that the symmetries are only approximate, but even if this was granted, the conclusion would be for a large number of
almost Pareto efficient states-each of which is almost certainly close to an actual Pareto efficient state than they are to each other. 

This means that symmetry breaking plays an important role in the the choice of allocation made by a market mechanism in any realistic model of an economy, in which there are similarly endowed households or similar products.  And this implies that equilibria,
assuming they exist, are highly non-unique.

\section{Critique of the Arrow-Debreu model of an economy}

We can sketch here some of the strengths and weaknesses of this model of economics.  Little that I have to say here will
be original.

\subsection{Strengths of Arrow-Debreu}

The major strength of the Arrow-Debreu formal model of an economy is its generality.  It makes some major and sweeping 
idealizations, which have been criticized extensively-and which we will discuss shortly. But within those limitations it is
remarkably general.  

When restricted to a simple economy with one or two households and companies it reproduces the simple reasoning based on
the Edgeworth box.  But it can be applied to economies which are arbitrarily large and complex.  There are several assumptions 
which are hard to criticize:  does someone want to propose that most companies do not choose to act to maximize their profits?
Or does someone want to propose that individuals don't try to get the best salary for their labour or that households do not decide
to buy what they prefer more to have\footnote{Of course it is possible to reply here that individuals have confused and inconsistent
preferences, and make irrational decisions.  But it is still a strength of the Arrow-Debreu model  to show that under the assumption that
people are rational, equilibria where all markets clear exist.  It then is an area of research to explore the implications of limits to
rationality. }?

In a small model with two goods and agents, one can show by hand the generic existence of Pareto efficient choices and 
the corresponding prices.  But it is not trivial to show that there are Pareto efficient configurations for very large and 
complex economies.  The Arrow-Debreu formulation accomplishes that.  Nor is it at all obvious that under very general
assumptions that there should exist choices of prices which allow every one to extremize their profits and preferences.

Related to this is what Eric Weinstein has called the {\it canonicalness} of this formulation of economics.  It is like
quantum mechanics or classical mechanics, in that it is not easily modified.  Attempts to modify it often come back to the
same formulation, with an adjustment of endowments or other inputs.  Or else they introduce features that lead to a much
different and less tractable model.  So like classical and quantum mechanics, this model of economics seems to occupy a unique
and isolated point in the space of theories, it is not one of a large set of similar models making slightly different assumptions.

Finally, a great achievement of the Arrow-Debreu approach are theorems that show that it can not be the basis of ideological 
viewpoints on economies such as the desirability of large or small differences in wealth or income or the role of government in regulation and or delivering services such as education and health care.  One can show that whatever distribution of endowments there are equilibria which are Pareto efficient. Since endowments could include education and basic services that make work possible, such as access to health care, as well as initial wealth, the fact that there are Pareto efficient equilibria for very wide ranges of choices of endowments suggest that the markets will function well under wide ranges of policies with respect to eduction, health care, taxation etc\footnote{This is of course a complicated issue, but to put the point a different way: is there a claim that economic and financial markets in europe are not
in equilibrium because health care and education are socialized?  It would seem that if the law of supply and demand suffices to find equilibria which are Pareto efficient in an economy where health care and education are privatized, they would function the same way in a mixed, social democratic economy.}.

\subsection{The treatment of space in Arrow Debreu}

There is no mention of space in the original Arrow-Debreu model.  But if there is any reason to distinguish goods produced or consumed at different places, these are just treated as different goods.  
One way to see this is that transportation of a book case from the factory in Kitchener to the store in New York is just another production process.  The input is the book case in Kitchener, the output is a book case in New York.  Given this, the basic assumptions and results of the model are unchanged.

\subsection{The issue of time in Arrow Debreu}

We can now come to the main issue which is the treatment of time in the Arrow-Debreu model of an economy.  It is apparent that
time is treated in a way analogous to one aspect of its treatment in Newtonian physics.  In classical mechanics, we say that time
has been {\it geometrized} or {\it spatialized} in that the time coordinate is treated just like one of the other coordinates.  Motion in 
a $d$ dimensional space is treated as a geometrical curve in a $d+1$ dimensional space.  

Following this, it is natural in Arrow-Debreu to treat time the way space is treated.  A single good at two times is treated as two 
distinct goods, just as a  goods at two places is treated as two goods.  The prices are treated as being set at one time, so that goods to be delivered in the future may have a different price today than the same good delivered today.  

Making use of this,  we can say that time, like space, is treated like just another feature of a commodity.  The essential formulation is
timeless, indeed, time plays no role and need never be mentioned.  In this case the model corresponds to a market which meets at a single time, at which all production processes and consumption plans are chosen. There is no dynamics under which the system approaches
equilibrium, equilibrium is in Arrow-Debreu just a timeless characterization of a distinguished state of the system. 

The point is that there is only a single market, which takes place at the initial time.  The endowment of the household include the earnings of all its members over all their lifetimes.   Similarly the production plan of each firm is now a plan over the life of the firm, decided in full at the beginning.  Similarly , each consumption plan sets out what will be made and consumed at every future time, and supply and demand are equated at every future time, leading to a single, timeless notion of equilibrium.   

There is no dynamics, no development, no learning and certainly no irreversibility.

\subsection{Treatment of contingency in Arrow Debreu}

The assumptions of the original Arrow-Debreu model make no mention of contingency or uncertainty, just as they make no mention of time or space.  There is a single list of possible goods, fixed for all time, and known to all participants.  In each state of the system each good has a single universal price.  

Just as time and place are incorporated by multiplying the set of goods, contingency is treated by introducing distinctions between 
the same good under different conditions.  For example, suppose it matters to prices next summer whether it rains or not next March.
Then, for each good next August, there are now two goods, one under the contingency it does rain, one under the contingency it does not.

The model is then extended by making some spectacular assumptions about the possible contingencies the market may be subject to.

\begin{itemize}

\item{} The contingencies have all to do with things apart from decisions made by participants in the market, or the state of the system.  They concern things like weather, discoveries of raw materials, new technologies etc.

\item{} The contingencies have the structure of a tree.  Each contingency is a path through that set from the
initial moment the market happens to conditions at a future moment,  Since the set of contingencies is a tree, each element implies its past
elements.  A contingency is also called a "state of the world".   

\item{}It is assumed that all possible contingencies for all future times in the life of the market are known\footnote{{\it "Uncertainty means we do not know what will happen in the future.  But we do know what might happen."} Starr \cite{Starr}, p 185}. 

\end{itemize}

Combining space, time and uncertainty, there is then a vast explosion in the number of goods.  Rather than having a particular model of
Ikea made chair, we have a vast set of chairs, identical except for the place, time and contingent state of the world.  Each of these
is called a {\it localized, dated, contingent good}.  

The basic model is then extended again.  Each production plan is a vector over the space of contingent dated goods.  It tells what the firm will produce from which inputs and outputs, at which times and under which circumstances. Similarly, each consumption plan is a precise
plan for the household over its whole life, subject to all the contingencies specified in the model.  The preference or utility function now
ranges over the whole set of dated contingent goods.

Under these assumptions the model extends trivially and the same results on the existence of equilibria prices and their Pareto efficiency can be shown.

\subsection{Weaknesses of the results of Arrow Debreu}

Let us postpone for a moment any criticism of the assumptions of the mode, and take them as given.   There are several obvious
weaknesses.  

\begin{itemize}

\item{} It is proven that equilibria exist, and they are Pareto efficient, but it appears that generically equilibria are far from unique.
 This means that one cannot argue that the market, by itself, can find the state of {\it maximal} effeciency or
utility.  There appear instead to be many possible attainable states which are all Pareto efficient.  This means that additional
criteria must be imposed by the society to decide which state is most desirable.  By definition these criteria are in addition to the
maximization of utility functions and profits.   It then appears that economics may suffer from a {\it landscape problem}, analogous to that
in string theory.  This would mean that the explanation for why one equilibria is realized in an economy rather than another cannot lie in the market forces modeled by the Arrow-Debreu model, because those market forces allow a large set of possible equilibria.  

\item{} There is no mechanism given for whether or how equilibria configurations are reached by the model.  Furthermore, since
equilibria are in general non-unique, there is no mechanism in the theory to explain why one rather than another of these
equilibria could be chosen by the market mechanism.  All we know about the market mechanism in the theory is that it
looks for Pareto efficient states, but if there are many the market mechanism cannot choose among them.  

\item{} There are other questions  we would like to ask, such as how the rate of convergence to equilibria depends on parameters
of models generically, such as number of goods firms and households.  There do not seem to be general results of this kind.  This is relevent because one might argue that the assumptions of the model are only valid for certain short times, before unpredictable
events or changes in technology intervene, and the predictions of the model are only interesting if equilibrium is reached in times short compared to these.

\item{} One way to express these points is that what is lacking is a non-equilibrium dynamical theory that explains how, when and how fast equilibrium is achieved.  In physics there is such a thing and it greatly deepens our understanding of thermodynamic equilibrium.  While there is not a close analogy between economic and thermodynamic equilibrium, it is nonetheless the case that the economic notion of equilibrium would be deepened by understanding the non-equilibrium dynamics that leads to equilibrium. 

\item{}  While there is an intuitive sense that the basic ideas underlying neoclassical economic models capture something true about markets, it would be interesting to know if there is a body of results that test the theory against detailed data.   To approach this we want to know far from  ideal a real economy may be, and still count as evidence for the theory. For example, lets take the prediction that all markets clear in equilibrium.  There are clearly lots of markets in the real world that do not perfectly clear.  An example is the book business: around half the books which are printed are never sold.  Is this a success of neoclassical economics, because the market comes within a factor of two of clearing, or a failure, because one can imagine markets that come much closer.  Indeed publishers have a simple motivation to cut costs by only printing the books that will sell, but it seems very difficult to predict accurately which books will sell and which won't.  One rule of thumb-to which there are exceptions-is that books that are not in book stores don't sell.  So one can imagine arguing that the physical book is partly an advertisement for itself-in which case 50 percent may be good. But the point is, how are we to tell?

Another aspect of this problem is that without a dynamical formulation its hard to know the time scale over which markets should clear in equilibrium.  There is an average time it takes a house to sell, when this turns into months rather than weeks is this is a sign the housing market is not in equilibrium?

\item{} One thing that an equilibrium model might be able to do is characterize  a phase of an economy when it is approaching equilibrium and differentiate it from one where equilibrium has been reached.  One would expect, for example, that in a financial market, a lot more trading goes on in the process that discovers and reaches equilibrium than afterwards, when the dynamics just maintains the state of equilibrium.  This issue is discussed in \cite{FG} where an interesting issue was raised: if stock and currency markets are in equilibrium then why are the average daily trading volumes so large?

\item{} If the equilibria are generically non-unique, they are generically unstable as well.  There do not appear to be general results
about stability of equilibria.  If they are unstable they are not useful.  One might imagine a kind of economic model where Pareto efficiency was associated generically with instability, so that maximally efficient states of the system were most often unstable, leading to large fluctuations in prices and big changes in consumption and production plans.  It might be that other states were stable or quasi stable, but at a cost of giving up some efficiency.  Suppose that an economy was much more stable in states only half as efficient as Pareto 
efficiency, in terms of profits and utilities.  Might not there be wisdom in choosing to be a bit less affluent but more stable?

\item{} It is sometimes claimed that one can deduce from this formulation that maximal utility (ie Pareto efficiency) is achieved by 
a pure laissez-faire  market economy, without intervention by the state\footnote{{\it "...On this basis, public authority intervention in the market through direct provision of services (housing, education, medical care, child care etc.) is an unnecessary escape
from market allocation mechanisms with their efficiency properties.  Public authority redistribution of income should be sufficient to
achieve the desired reallocation of welfare while retaining the market discipline for efficient resource utilization.}. Starr \cite{Starr} p 151}.  
However, if there are many Pareto efficient states of an economy such an argument cannot be made.  Moreoever, in a 
mixed economy, government provision of such services can be modeled as external constraints on accessible production and
consumption plans, and the results on existence of equilibria and their Pareto efficiency should still be provable.  

\end{itemize}

\subsection{Weaknesses of the assumptions of Arrow Debreu}

However, the situation is worse than just described, because there is much to criticize about the applicability of the 
Arrow-Debreu model to reality based on its assumptions.

\begin{itemize}

\item{} The idea that there is a fixed set of dated, contingent, goods, known to all participants, fixed over time is false.  In the modern economy goods are often introduced which quickly come to dominate markets that were not previously known about or anticipated.  

\item{} The idea that all contingencies are known in advance is also highly unrealistic.  Lots of things happen that are not planned for specifically.  Few individuals are able to predict their lifetime schedule of earnings within an order of magnitude, and indeed, very few
households plan for more than a few years into the future, at most.  Few couples can predict how many children they will have and when they will be born.  No one is able to predict what their children will be interested in.  Imagine asking a couple on their marriage to budget all the vacations they will take for the rest of their lives, or to give preferences amongst all the opportunities they may have to travel, to take up new hobbies, or to consume products not yet invented.  

For example, consider the adage, well known to sailors, {\it "The two happiest days in the life of a boat owner are the day they buy their boat and the day they sell it."}  Imagine coding this in a utility function. Without contingency it is impossible to do it, but a little thought will convince the reader that coding this into a vast tree of possible contingencies is going to be awkward at best.   The reality is that the essence of human life is we learn and grow as we live, and a great deal of it necessarily comes as a surprise.  

\item{} This means that preferences change, due to changing circumstances, new unanticipated opportunities as well as changing tastes.  The inability to deal with this in neoclassical economics is well known and has a name, which is the {\it changing preferences problem.}  It seems likely this is a key problem that goes to the core of the model's assumptions and mathematical expression, and signals major revision in the theory.  

\item{}Similarly, few firms plan successfully more than a few years in advance, and few consider the whole tree of known contingencies even within that window.  

\item{} There is an  enormous combinatorial explosion in numbers of distinct goods that the treatment of time and contingency implies must exist.   It is trivially true that, given the assumptions of their model, the results still obtain even given vast combinatorial increases in the numbers of goods.  But if the model is to be relevent there must be markets whose dynamics discover and set prices for this vast array of goods.   As pointed out by Starr, \cite{Starr}, p 193-4 in the real world there are no markets which consider the vast number of dated
and contingent goods required for the Arrow-Debreu model to be applicable.   Given this, the extension of Arrow-Debreu to incorporate time and contingency seems to make assumptions that are far from realized in real economies.  But if the model is not so extended it is even less correct as a model of real markets.  

\item{} The bottom line is that real markets are often significantly perturbed by developments which were not previously anticipated.  Moreoever real economic decision makers of households do not have preference functions that encompass a long list of possible dated contingent goods.  They have a few things they need and a slightly longer list of things they want.  I would guess that for most households, the list of what they need and want at a given time is not longer than can be printed in a few pages and is at least in part ever changing.  
Most households have no preference at all over vast numbers of choices, because they represent things they either have no interest in or have no reason to expect they would be able to purchase.  

There seems to be no reason except mathematical formalism to assume that each household and firm deals with decisions concerning the same set of goods.  Functions on $R^n$ is a nice mathematical setting, but it should not force us into silly idealizations that make the model
inapplicable.  Why not construct a model in which each household has a short list of goods they have acquired, have for sale or would be interested in acquiring?

Similarly, real firms specialize in a small part of the market, they do not need to have profits expressed as functions over the
whole vector space of possible inventories.  More significantly, most firms do not plan outside of a small window of time and future contingencies, because they know from experience this is futile.  Instead, in the real economy firms, like households, are ever evolving and ever learning.  They make the decisions that face them at the time they need to be made.  All decisions are made with vastly incomplete knowledge of the full situation and full implications.  A realistic model of an economy would be based on agents whose knowledge is
strongly limited.  

\item{} Finally, there is an implicit assumption that all decisions that affect a market take the form of choices of consumption or consumption plans.  But this is clearly not the case.  In real life households have preferences which cannot be expressed by maximizing individual utility functions.  People have diverse preferences about how important it is for them to live in a society where everyone has access to good education or health care or the extent to which financial markets are regulated to improve their stability.  These preferences are expressed by voting and, to a significant degree, by choosing where to live and work.  As we saw, the fact that there are multiple equilibria means that there must be forces outside the market which choose one of many possible equilibria, it is interesting to ask whether there might be an extension of Arrow-Debreu where these are also modeled\cite{sabine}.

\end{itemize}

\section{Taking time into account}

The Arrow-Debreu model is very impressive, within the range of its actual applicability.  Let us see what lessons we can draw
to go beyond it.  What is clear is that there is a need for a non-equilibrium theory of markets that would allow us to understand questions that are not accessible from the framework of the Arrow-Debreu model.  An analogy to physics might be helpful here.  Just like there is micro and macro economics, there is micro and macro physics.  The former is atomic physics, the latter includes thermodynamics and the description of bulk matter in different phases.  Macrophysics mostly deals with matter in equilibrium.  The bridge between them is a subject called {\it statistical mechanics}, which is a general study of the behavior of large numbers of atoms, both in and out of equilibrium.  Indeed, even though there is not a close analogy between the notions of equilibrium in economics and physics, there is clearly a need for a subject
that might be called {\it statistical economics}.  It would be based on a microscopic model of the basic agents and operations or processes that make up an economy and study the behavoir of large numbers of them in interaction.  Some of the questions that might be 
addressed by such a formulation would be:

\begin{itemize}

\item{} Derive a notion of steady states for markets and show that to some approximation they correspond with the notion of equilibrium in neoclassical economics.  Different equilibria or steady states might then correspond to different phases of the models.  

\item{}Study the phase structure of economic models, and determine what are the important macroscopic observables that can be tuned to choose different phases.  

\item{} By doing so, derive some of the phenomenological laws or observations of macro economics, such as the law of comparative advantage.

\item{}Study the passage to equilibria from generic out of equilibrium states, and determine how the relaxation time to equilibrium can depend on macroscopic parameters.

\item{}Study fluctuations around equilibrium, and in particular characterize whether they are Gaussian or power law distributed in different kinds of markets.

\end{itemize}

To summarize, just as in physics, even if the system we are interested in is in equilibrium most of the time, there is a need for a non-equilibrium theory of economics, to deepen our understanding of the state of equilibrium.

\subsection{Some observations and candidate principles for non-equilibrium economics}

We begin the search for a non-equilibrium economic theory, by taking the lessons we have learned from the successes and failures of Arrow-Debreu and seeing if we can turn them
into useful principles to guide the construction of better models.  The key issue as we have mentioned before is to be more realistic about the nature of time and the role of uncertainty.

\begin{enumerate}

 \item{}{\it \bf Time must be incorporated in a way that recognizes the irreversibility of most actions taken, as well as the asymmetry of the present, past and future.}   An economic theory must be based on decisions made by individual agents at a given time, on the basis of records they hold at that moment, of the past.

\item{} {\it \bf The future is uncertain,}  and no amount of mathematical precision can make economics into a tool that could give unique predictions for 
the specific evolution of large, complex economies.   This is for two reasons.  The space of known contingencies and possibilities explodes combinatorially so that there is no possibility of representing it usefully.  Second, are the Taleb's Black Swans\cite{blackswans}, ie all the events that will not be predicted no matter how wise or careful we are.
 
 Still economics can be a science.  For example, it would be quite sufficient if, like biology, it tried less to predict the future and tried more to explain the present and past.  This is a much easier problem, because as we go back into the past we the space of contingent possibilities narrows rather than grows.  We don't know what will happen but we know a great deal about what has happened.  This is because we have a great deal of information available in the form of accounting records, and other records.  
 
\item{}  {\it \bf The observables of economics are accounting and other records. }
 
One should then try to construct a theory of economics that involves only observables. The importance of this kind of operational principle in physics and other sciences have been paramount.  Let's see what it can do for economics.  
This restricts attention to what is actually measured by companies, individuals and governments. And it removes from consideration fictional elements that have nothing to do with how real economies work such as fixed spaces of products, fixed production plans, utility functions etc.   
 
 \item{} {\it \bf  Agents in economic systems make decisions at each moment of time on the basis of  information contained in their own records and public information.}  This information is always incomplete, and in particular, is commonly insufficient to force a unique outcome of decisions as to what to do in the next time step.  There is no need to posit perfect information or perfect rationality.
 
  \item{\bf Thermodynamic equilibrium is not analogous to physical equilibrium because an economic system is not isolated.}  An economy is an open system which has inputs in the form of human work, energy, raw materials and the invention of new technologies and products.  Therefor economic equilibrium may be analogous to the kind of {\it non-equiibrium steady states} that occur in such open systems in physics.  If this is the case then relevant observables include rates of flow of critical materials through the system and around closed cycles in the system\footnote{It is interesting to wonder whether the cycle theorem of Morowit\cite{morowitz} is relevant. This asserts that
cycles necessarily form in a network of reactions that comes to a steady state.}.  
 
 \item{}{\it \bf Typical economies may have access to a large number of possible,
 quasi stable steady states.  Moreover,  to the extent that these can be approximately characterized as equilibria, they exist and are
 efficient under large classes of choices of endowments and constraints. }{\it  This suggests that decisions about allocation of wealth, education and other endowments must be made on the basis of ethics and politics endogenous to any formal model of a market economy.}
 
 \item{}{\it \bf Markets with large numbers of agents have very large approximate symmetries, expressing the fact that there are many 
 individuals with similar educations, interests or aspirations and many firms competing to offer similar products and services.  In the steady states reached by real economies these symmetries are usually broken.}  This leads to multiple equilibria or steady states.  The representation theory of the broken symmetries is then relevant to the distribution of equilibria.  
 
 \item{}{\it \bf   There are gauge symmetries associated with rescalings  of the units used to
 value individual goods.  The dynamics of markets should be invariant under these gauge transformations.  } This will be discussed in detail in the next section

\end{enumerate}

\subsection{Agent centric models of economic markets}

To study and test these candidate principles one can construct models.  Given the tools available now, that were not available to earlier generations of researchers, one can learn quite a lot just by building a computer model in which a large number of agents interact and study it.  
A large number of agent based models of markets have been studied and it appears that they provide the  best way to approach the study of non-equilibrium, or statistical economics.  

Among the models which have been constructed, a class which inspired the ideas discussed here are the  partecon models, developed by Brown, Herriot, Kauffman, Palmrose and Sawhill\cite{partecon}.  Two agent based models along these lines have been built 
and studied, by Jim Herriot\cite{parteconI}, and  Samuel Vazquez\cite{parteconII}. 
 One of the key ideas in these models is to represent companies with data structures derived from accounting principles.  
 The agents trade bilaterally and the information each agent has is only what they have learned in their prior history of trading.
So there is no global price, instead each agent stores the history of successful and failed bids and asks.  Furthermore, there is not necessarily a single currency,two agents may trade any elements in their inventory.  It is then possible to study the emergence of a single good, or representative of a good, which becomes money.  One can also study many other things such as whether, under what circumstances, and how fast the community of agents converge to a common price for a particular good, and how large are the fluctuations around that common price.  More about these models is discussed in \cite{parteconI, parteconII}

Generalizing from these models, one can define an agent based economic model in the following way. 

\begin{itemize}

\item{}  {\it Goods} are material goods or services, capable of being owned, transformed and traded.
\item{} {\it An economic agent} is a model of a legally recognized person, family  or corporation capable of the following:

\begin{enumerate}
\item{} They own stuff, which they keep an inventory of.
\item{}  They may transform stuff to other stuff.
\item{}  They may trade stuff.
\item{} 	They may make contracts for trades at later times.
\item{} They have needs, which must be fulfilled if they are to survive as well as aims or goals.  
\item{} 	They make decisions about the forgoing
\item{} 	They keep records of the forgoing. 
\item{} 	In doing so they are governed by externally imposed rules and laws.

\end{enumerate}

\item{} An {\it economic operation} is a change in the state of one or more agents by virtue of one of the following kinds of events:  
\begin{enumerate}
\item{} A process, or transformation in one agent is a change of some stuff in the inventories of a single agent. 

\item{} A trade, in which two agents exchange two goods.  
\item{}  There may be processes in which agents are born or die, and in which new products are introduced to the system or inactive products removed from the system. 
\end{enumerate}
\end{itemize}

The claim underlying this kind of modeling is that all the dynamics of an economy is made up of these kinds of economic processes.  The observables of economics are then nothing but the records of the inventories of the agents and of these processes.  

An economic model is then specified by giving 

\begin{enumerate} 
\item{} 	a list of models of  N agents, or an algorithm to produce them (including possibilities for birth and death of agents).
\item{} 	P kinds of goods, which may be traded, owned and  transformed.  
\item{} 	strategies for decision making of the diverse agents.
\item{} 	External conditions driving the system, such as inputs of energy, materials, innovations, etc. and outputs for waste.

\end{enumerate} 

Given all this one can make a stab at a definition of statistical economics:

\begin{itemize}

\item{} {\bf  Statistical economics is the study of the collective behavior of large numbers of economic agents.  }
	
\end{itemize}

\section{Gauge invariances in economics}

We can now  turn finally to the role of gauge invariance in economics.  The proposal that gauge theory is essential for making progress in economics was made by Malaney and Weinstein\cite{MW}.   I would now like to argue that there are deep and compelling reasons why the extension of economic theory to incorporate out of equilibrium behavior should centrally involve the kinds of gauge invariances they proposed.  

\subsection{Why gauge invariance in economics?}

As we saw, the Arrow-Debreu model itself has gauge invariances, described by  (\ref{scalingp}) and 
(\ref{scalingU})\footnote{While this is an elementary observation, to my knowledge this has not been pointed out before.}.   The proposal of Malaney and Weinstein is that to construct models of economies that have real dynamics and time dependence in them- so that for example, preferences of households can change in time-it is necessary to hypothesize that the dynamics is constrained by much larger groups of
gauge invariances.  

As we have seen in the discussion above, the need for gauge invariance  stems from a fundamental fact about prices, which is that they appear to be at least in part arbitrary.  It seems that each agent in an economic system is free to put any value they like on any object or commodity subject to trade.  How do we describe dynamics of a market given all this freedom?  To get started we recall that 
in the Arrow-Debreu description of economic equilibrium, there is a gauge symmetry corresponding to scaling all prices,
(\ref{scalingp}).  This may suffice for equilibrium, but it is insufficient for describing the dynamics out of equilibrium, because away from equilibrium there may be no agreement as to what the prices are.  There is then not one price, but many views as to what prices should be.  Each agent should then be  free to value and measure currency and goods in any units they like-and this should still not change the dynamics of the market.   It should not even matter if two agents trading with each other use different units. Thus we require an extension of the gauge symmetry in which the freedom (\ref{scalingp}) is given to each agent, so they may each scale their units of prices as they wish, independently of the others.  

There is a further difficulty with price which is that even after issues of measurement units are accounted for different agents will value different currencies or goods differently.  Different agents have different views of the economy or market they are in, they have diverse experiences, strategies and goals, and consequently have different views of the values of currencies, goods and financial instruments.   Consequently, in a given economy or market it is often possible to participate in a cycle of trading of currencies, goods or instruments and make a profit or a loss, without anything actually having been produced or manufactured.  This is called arbitrage.  

In equilibrium, all the inconsistencies in pricing are hypothesized to vanish.  This is what is called the {\it no arbitrage} assumption.  But  out of  equilibrium  there will exist generically inconsistencies in pricing.   In fact, we are very interested in the dynamics of these inconsistencies because we want to understand how market forces act out of equilibrium on inconsistencies and differences in prices to force them to vanish.  This is essential to answer the questions the static notion of equilibrium in the Arrow-Debreu model does not address.  

However, in analyzing the dynamics that results from the inconsistencies,  we need to be careful to untangle meaningful differences  and inconsistencies in prices from the freedom each agent has to rescale the units and currencies in which those prices are expressed.  This is precisely what the technology of gauge theories does for economics.  

How does gauge theory accomplish this?  As in applications of gauge theory to particle physics and gravitation, the key is to ask what quantities are meaningful and observable, once the freedom to rescale and redefine units of measure are taken into account.  
The answer is that out of equilibrium  the meaningful observables are not defined at a single event, trade or agent. 
Because  of the freedom each agent has to rescale units and choose different currencies,  the ratios of  pairs of numerical prices held by two agents in a single trade are not directly meaningful. 

To define a meaningful observable for an economic system one must compare ratios of prices of several goods of one agent, or consider the return, relative to doing nothing, to an agent of participating in a cycle of trading.  This might be a cycle of trades that starts in one currency, goes through several currencies or goods and ends up back in the initial currency.  Because the starting and ending currencies are the same, their ratio is meaningful and invariant under rescalings of the currency's value.  This is true whether one agent or several are involved in the cycle of trades.  We say that these kinds of quantities are {\it gauge invariant.}

Such quantities, defined by cycles of trades such that they end up taking the ratio of two prices held by the same agent in the same currency have a name: they are called curvatures.  The ratios of prices given to a good by different agents also have a name: they are called connections.  The latter do depend on units and hence are gauge dependent, the former are invariant under arbitrary 
scalings of units by each agent and are gauge invariant or gauge covariant (this means they transform in simple fashion under
the gauge transformations.)

It is interesting that the quantities that are invariant under the gauge transformations include arbitrages, which should vanish in equilibrium. This does not mean that they are irrelevant, indeed, they may be precisely the quantities one needs to understand how the
non-equilibrium dynamics drives the system to equilibrium.   That is, it is natural to frame the non-equilibrium dynamics in terms of quantities that vanish in equilibrium.  These are the quantities that the law of supply and demand acts on, in order to diminish them.  

There is a precise analogy to how gauge invariance works in physics.   In gauge theories in physics, local observables are not defined because of the freedom to redefine units of measure from place to place and time to time.  Instead, observables are defined by carrying some object around a closed path and comparing it with a copy of its configuration left at the starting point.  These observables are called  curvatures.  The result of carrying something on a segment of open path is called a connection and is dependent on local units of measure. But  when one closes the path, one makes comparison to the starting point possible, so one gets a meaningful observable, which is a curvature.  

In general relativity exactly the same thing is true.  Here curvature corresponds to inconsistencies in measurements, for example, you can carry a ruler around a closed path and it comes back pointing in a different direction from its start.  The dynamics is then given by the Einstein equations, which are expressed as equations in the curvature.  But in the ground state-which is roughly analogous to equilibrium in an economic model-the curvature vanishes.  The state with no curvature, called flat spacetime, is the geometry of spacetime in the absence of matter or gravitational forces.  It is the state where all observers agree on measurements, such as which rulers are parallel to which.  

But while the curvatures vanish in the ground state, the physics of that state is best understood in terms of the curvatures. 
For example, suppose one perturbs flat spacetime a little bit.  The result are small ripples of curvature that propagate at the speed of light.  These are gravitational waves.  The stability of flat spacetime is explained by the fact that these ripples in curvature require energy.

Similarly, it may be that the stability of economic equilibrium can be studied by modeling the dynamics of small departures from equilibrium.  These are states where prices are inconsistent, ie where arbitrage or curvature is possible.    By postulating that the dynamics is governed by a law that says that inconsistencies evolve, one gets a completely different understanding of the underlying dynamics than in a theory that simply says that inconsistencies or curvatures vanish. 

If economics follows the model from physics, then the next step, after  one has answered the question of what are the observables, is to ask what are the forms of the laws that govern the dynamics of those observables. Given what we have said, the choice of possible laws is governed by a simple principle: {\it since only gauge invariant quantities are meaningful, the dynamics must be constructed in terms of them alone.}  This principle is highly constraining in physics.  For example, in  general relativity and Yang-Mills theory it leads, to a given order of approximation, to a unique dynamics.    
If the same holds in economics, then the study of non-equilibrium markets may be much more constrained than one might have otherwise thought. 

Below we will begin the study of such models.  
Vazquez in \cite{parteconII} has built and run several agent based models of markets where the dynamics is constrained by a principle of gauge invariance, and found a number of interesting phenomena.  He is able to study how a market approaches equilibrium, and then fluctuates around it.  These are encouraging first results, but much remains in this direction to do\footnote{The paradigm of gauge theories in physics and mathematics can be approached from different points of view, and likely the same is true in economics as well.  It can for example, be approached from a viewpoint of the history and philosophy of physics, within which the 
setting is the discussion of whether space and time are absolute or relational.   Quantities are absolute when they can be applied to one entity alone, without reference to whether it is isolated or part of a larger system. They are relational when they describe relationships between entities that interact with each other.  Physics made the transition from an absolute-Newtonian-way of understanding space and time to a relational view with the invention of Einstein's theory of general relativity.  Classical economics appears to be  based on an absolute framework in which prices given by one trader to one good are meaningful, whether that trader lives on an island with one other trader or is part of a diverse and complex economy.  This may suffice to characterize equilibrium but to describe the dynamics out of equilibrium we must model the trades between pairs of agents as they disagree over prices, and this requires a relational framework.   Agent based models such as the particon model are relational in the sense that prices emerge from a complex system when it dynamically reaches a steady state.  Gauge fields are the right language to model relational systems in physics, and they will be in economics as well.  

There is also a formal, mathematical argument for gauge invariance in economics.  This begins by asking what is the right way to deal with time and contingency.  In the neoclassical theory described above one takes the direct product of the sets of times or contingencies with the vector space of goods, so that time and contingency are treated just as different labels of products, on the same level as any other.  This neglects the fact that time and contingency have very particular structures.  The right way to deal with them is with mathematics that recognizes the ways in which time and contingency differ from other properties.  This is, for the case of time, to construct a vector bundle over the real line.  The real line represents time, the fibers are the spaces of products.  For the case of contingency, the tree of contingent states of the world is a special case of a partially ordered set and one can use a construction from category theory which is called a pre-sheaf in which one associates a vector space with each set of contingencies.  In either case, one has to ask how observables are to be coded into the description. The answer that naturally arises from the mathematics is gauge fields.  

In both cases, the mathematics tells us that we need structures to tie together measurements made at different times or under different contingencies.  This is formally where gauge degrees of freedom arise.}.

\subsection{The Malaney-Weinstein connection in economics}

There is no better illustration of the ideas just discussed than the original proposal of Malaney and Weinstein for introducing gauge invariance and gauge connections in economics\cite{MW}.  This was first described in Maleny's 1996 thesis, partly based on joint work with Weinstein.  Here is a brief synopsis of one of the results\footnote{Note that the notation used in this section is standard in physics, but may
not be known to economists.  In \cite{MW} these results are stated in notation more standard for economics.}. 
 
Consider an {\it economic history}, which is a curve in $ \alpha (t)$ in ${\cal  P} \times{\cal P}^*$.  It gives a sequence of inventories and prices.  For simplicity we will assume $t$ runs from
$t=0$ to $t=1$.  It will be convenient to assume that the total value of the goods $p_a q^a$ never vanishes so let us define $\cal C$ to be the subspace of ${\cal  P} \times{\cal P}^*$ such that $p_a q^a=0 $ and let the curve $\alpha (t)$ live in ${\cal R}= {\cal  P} \times{\cal P}^* -  {\cal C}$.  
 
Now consider the problem of computing a cost of living correction under the assumption that both prices and a standard baskets of goods change in time. This is necessary as changing technologies, fashions and products means that the basket of goods used to measure the consumer price index cannot be frozen. 
$\alpha (t)= (q^a (t) , p_b (t)) $ is then the time changing basket of goods, $q^a (t) $, and the changing prices, $p_b (t)$. 
 
Malaney gives a simple and elegant solution to this problem. The point is that there should be no increase or decrease in the cost of living resulting from a change in the standard basket as long as the change involves a swap of goods that have the same price at the time the swap is made.  
 
To compute the real change in the cost of living construct an Abelian connection on $\cal R$,
 \f
A=\frac{q^a dp_a}{ q^c p_c }
 \ff
We will call this the Malaney-Weinstein connection. 
Under a global rescaling of prices (\ref{scalingp}), possibly time dependent, $p_a \rightarrow \Lambda p_a$
 \f
A \rightarrow A + d \ln (\Lambda )
 \ff
So A is a connection for a global gauge symmetry of changing prices. Then the change in the cost of living is given by  $P$, an abelian holonomy 
 \f
P = e^{\int_\alpha A }
\label{Pdef}
 \ff
along the curve $\alpha (t) $.  
 
The point is that if the tangent to the curve alpha is in a direction that constitutes barter, ie no change in value at that $p_a$ and $q^a$, the contribution to $P$ vanishes. Malany shows that this is equal to the Divisia index\cite{divisia}  and that it has better properties than indices that involve simple ratios of costs of baskets of goods at two times. From our point of view this makes sense as $P$ transforms gauge covariantly under time dependent rescalings of currencies 
\f
P \rightarrow \frac{\Lambda (1)}{\Lambda (0)} P
\label{covariant}
\ff

Here is another problem the Malaney-Weinstein connection might be applied to.  Suppose that $\alpha (t)$ represents the
inventory and prices of a single agent during a sequence of trades of goods in which they participate.  The agent wants to compute the total profit or loss, in their currency, of the sequence of trading.  The profit or loss should transform covariantly under rescaling of 
their units of currencies, ie as (\ref{covariant}).  A proposal for an answer is again $P$ given by (\ref{Pdef}).

If one wants a completely gauge invariant quantity one has to take an economic history which is a closed curve, ie starts and ends at the same inventory and the same set of prices.  In this case from (\ref{covariant}) we see that $P$ is invariant under gauge transformations.

Can one make a profit or loss from participating in such a cycle of trades?  The answer is yes, if the resulting $P$ around the 
economic history $\alpha $ is not equal to unity.  Of course it is up to the market whether one can find other agents to trade with who
will allow this outcome, the no arbitrage assumption is that when markets are in equilibrium one cannot find such opportunities.

We can use a standard mathematical result to connect this to the talk of curvatures above.  Consider an economic history that is a small
loop that 
encloses a small plaquette around a particular $(q^a,p_a)$ specified by small changes  $(dq^a_0 , dp_b^0 )$.  Then 
\f
P \approx e^F 
\ff
where $F$ is the {\it curvature} given by\footnote{For notation see any textbook on differential forms. $F$ is a {\it two-form}, analogous
to the electromagnetic fields.} 
\f
F= \frac{1}{q^cp_c} \left [  \delta^a_b -\frac{q^a p_b}{q^d p_d}
\right ] dq^b_0 \wedge dp_a^0
\ff
Note that $F$ is gauge invariant.  Thus, we see for example, that for arbitrage to be possible there must be a cycle of trades that involves both changes in prices and
changes of inventory.  

More details and results are in \cite{MW}.

\subsection{Gauge invariance in agent based models}

We have seen that two tools unavailable to Arrow and Debreu may be useful for extending their work beyond the notions of
static equilibria developed there.  The first is the use of agent based models to efficiently explore the behavior of large collections
of economic agents.  The second is the notion of gauge invariance, as first proposed by Malaney and Weinstein, we have just discussed.  
I now sketch a framework for combining them.

In this particular realization we take the point of view of the partecon models\cite{partecon,parteconI,parteconII}.  These are models where to start off with there is not necessarily a single currency.  
We start with a more general notion, which is exchanging arbitrary sets of goods for
each other.  The reason to do this is to try to understand the existence of money and prices as not given, but representing a solution
to a problem of making efficient markets, which the market discovers and implements.  From this more general point of view, a price is a ratio of a good to a currency, and one cannot understand what it is if only goods have dynamics. Both goods and currencies get their values from dynamics and to understand what their ratio means,  one must go back to the beginning and model a network of agents bartering with each other, and see how, dynamically, currencies arise, emerging from the complex dynamics of the agents.  

\subsubsection*{The description of a single agent}

The basic mathematical structure that characterizes a single agent is as follows:

\begin{itemize}

\item{}{\bf Time} All quantities are functions of and evolve in a discrete time, $n$.  

\item{}{\bf  Inventory.}  There are $P$ agents, labeled by $i,j=1,...P$ and $N$ goods, labeled by
$a,b,c...= 1,..., N$.   The state of inventory is given by vectors $V_i^a$ which describes the quantity of goods $a$ held by agent $i$.

\item{}{\bf Gauge principle:}  {\it  Nothing in the dynamics of an economic system  can depend on the units in which the inventories of the different agents are valued.}   Moreover the different agents are free to use different units or measures to value their own inventories.

Hence the observables and dynamics of an economy should be invariant under transformations
\f
V_i^a \rightarrow V_i^{a \prime} = \phi_i^a  V_i^a  
\label{gauge}
\ff
where $\phi_i^a \in R^+$, the positive real numbers.   Hence the gauge group is abelian, and is 
$(R^+)^{NP}$.  

\item{} {\bf Adjoint.}  The natural adjoint for a vector space valued in $(R^+)^n$ is inverse:
\f
*: V_i^a \rightarrow (V_i^a)^*= \frac{1}{V_i^a}
\label{adjoint}
\ff
\item{}{\bf Invariant norm:} This implies that there is a norm of vectors
\f
|V|^2 \equiv \sum_{i,a} (V_i^a)^* V_i^a = n=NP
\ff
This norm is obviously preserved by the gauge transformations (\ref{gauge}). 

\item{}{\bf What by what matrices} These represent the views of each agent as to the ratios of values of the different goods.  So this is a matrix for each agent: $W_{i a}^b$ is the value that the $i$'th agent would give on a trade
of $a$ for $b$; i.e. $W_{i a}^b= a/b$  That is, if $W_{i a}^b=3 $ it means that agent $i$ would trade three $a$'s for one $b$.

\item{}{\bf Gauge fields:} {\it The $W_{i a}^b$  are formally connections}, in that under the rescaling of units (\ref{gauge}) they transform as 
\f
W_{i a}^b \rightarrow W_{i a}^{b \prime}= (\phi_i^{a})^{ -1}W_{i a}^b \phi_i^b
\ff

\item{}{\bf Valuation of inventory}  From $W_{i a}^b$ and $V_i^a$ one may construct
\f
{\cal I}^b= W_{i a}^b V^a
\ff
This is a vector of values of the total inventory in units of each kind of stuff.  that is ${\cal I}^{15}$ is the value of the total inventory of $A_i$ in units of the $15$'th kind of stuff.  

\item{}{\bf Properties of what by what matrices:}

Here are three convenient definitions:

\begin{quote}

If agent $i$ has no view as to the ratio of values of goods $a$ and $b$ then we write
$W_{i a}^b=?$.   

\end{quote}

\begin{quote}

A what by what matrix $W_{i a}^b$ is {\it complete} if there are no $?$'s in the entries, so the agent has an hypothesis about all possible trades.

\end{quote}

\begin{quote}

A what by what matrix $W_{i a}^b$ is {\it consistent} if 
$W_{i a}^b= \frac{1}{W_{i b}^a}$ and
$W_{i a}^b = W_{i a}^c W_{i c}^b$ for all $a,b,c$.  

\end{quote}

If an agent's $W_{I a}^b$ is consistent and complete\footnote{Matrices which are consistent and complete are also known as {\it reciprocal matrices.}  Thanks to Simone Severini for this information.} than it is a proportional to a projection operator.  To see this note that consistency and completeness implies that there is a single currency, lets call it $a=0$ such that all goods have consistent prices in that currency, given by 
\f
P_i^a= W_{i 0}^a
\ff
We can then decompose
\f
W_{i a}^b = \frac{P_i^b}{P_i^a} = (P_i^a)^*P_i^b
\ff
We can then check that
\f
W_{i a}^b W_{i b}^c = \frac{P_i^b}{P_i^a} \frac{P_i^c}{P_i^b} = N  \frac{P_i^c}{P_i^a}= N W_{i a}^c
\ff

\end{itemize}

\subsubsection*{Economic operations}

The basic idea of the model is that an economic system evolves through a series of trades, which may also be called ``economic operations."  These involve a pair of agents.  the description of the trades has to take into account the gauge principle that holds that different agents are free to value goods in whatever units they like. 

In a basic economic operation, $n^b_j$ units of good $b$, valued in units of agent $j$ are traded to agent $j$ by agent $i$ while, in return, simultaneously $n^a_i$ units of good $a$, valued in units of agent $i$ are traded to agent $i$ by agent $j$.  We will be interested in their ratio which is defined as
\f
{\cal O}_{ia}^{jb}= \frac{n^b_j}{n^a_i}
\ff
This transforms as a connection under the gauge transformations (\ref{gauge})
\f
{\cal O}_{ia}^{jb}  \rightarrow ({\cal O}_{ia}^{jb})^\prime = \phi^{a -1}_i {\cal O}_{ia}^{jb} \phi^b_j
\ff

A related quantity that we will find useful is 
\f
{\cal Q}_{ia}^{jb}= \frac{n^b_i}{n^a_i}
\ff
This is the ratio of the amount of goods b traded from $i$ to $j$ to the amount of good $a$ traded back from $j$ to $i$, but in this case, both valued by agent $i$.  This transforms the same way the what by what matrices transform, which is to say locally at agent, $i$. 
\f
{\cal Q}_{ia}^{jb}  \rightarrow ({\cal Q}_{ia}^{jb})^\prime = \phi^{a -1}_i {\cal Q}_{ia}^{jb} \phi^b_i
\ff

There are also production processes in which some combinations of goods are transformed into other combinations of goods.  An agent
may be endowed with such a process, in which case it follows rules that tell it when to run them\cite{partecon,parteconI,parteconII}. 
A typical rule might be that an agent run its production process as often as it can ie as soon as it has in its inventory the inputs needed.

\subsubsection*{Curvatures and observables}

The gauge principle tells us that all observables of an economic system must be gauge invariant, which means they do not depend on the units chosen to value goods in.  These are characterized by curvatures which measure gains and losses in cycles of trading.  This is as in physics, where the gauge invariant observeables are constructed from curvatures.  

For example, let us  suppose we have a circle of trades involving three agents $i,j$ and $k$.  We can define the consequences as a curvature:
\f
{\cal R}_{ijka}^{\ \ \ d} \equiv {\cal O}_{ia}^{jb}{\cal O}_{jb}^{kc}{\cal O}_{kc}^{id}
\ff
This transforms under a gauge transformation locally at the agent $i$.
\f
{\cal R}_{ijka}^{\ \ \ d}  \rightarrow ({\cal R}_{ijka}^{\ \ \ d} )^\prime =  \phi^{a -1}_i {\cal R}_{ijka}^{\ \ \ d} \phi^d_i
\ff
So let us consider a diagonal element, ${\cal R}_{ijka}^{\ \ \ a}$.  This is gauge invariant and hence an obserable.  It specifies the ratio of good $a$ that was returned to $i$ from $k$ to that traded away from $i$ to $j$, hence it specifies the profit or loss of the total cycle, so far as good $a$ is concerned.

A different kind of curvature can be constructed from the ${\cal Q}_{ia}^{jb}$.  It is given by 
\f
{\cal S}_{ijka}^{\ \ \ d} \equiv {\cal Q}_{ia}^{jb}{\cal Q}_{jb}^{kc}{\cal Q}_{kc}^{id}
\ff
This also transforms under a gauge transformation locally at the agent $i$.
\f
{\cal S}_{ijka}^{\ \ \ d}  \rightarrow ({\cal S}_{ijka}^{\ \ \ d} )^\prime =  \phi^{a -1}_i {\cal S}_{ijka}^{\ \ \ d} \phi^d_i
\ff
so that its diagonal elements are gauge invariant observables.  

As we discussed above, these curvatures represent departures from equilibrium.

\subsubsection*{Effective Dynamics}

We want to describe effective dynamics for our economic system whose kinematics we have just given.  
The actual dynamics is given by the evolution under the trading rules of the agents in the agent-based model.  The effective dynamics is not this, it is an hypothesis as to quantities that are minimized or extremized  by the system of agents when they achieve a non-equilibrium steady state.  

To describe the steady state then we want to construct an action $S$, which is to be minimized.   There are two kinds of terms that may appear:

\subsubsection*{Bilateral dynamics}

The dynamics may be dominated by pairs trading.  In this case the  effective action is given by sums of pairs of agents, and measures the economic operations between the pair.  A simple such action is
\f
S^{trade}= \sum_{i \neq j} Tr \left ( W_i {\cal O}_i^{\ \ j} W_j {\cal O}_j^{\ \ i}  \right )
\label{bilateral}
\ff

\subsubsection*{Trading dynamics dominated by cycles}

Motivated by Morowitzes cycle theorem\cite{morowitz}  and the general observation of the importance of cycles in non-equilibrium systems we hypothesize that modern economies are dominated by cycles of trading.  Consequently we expect their dynamics to be dominated by terms of the following form
\f
S^{cycles} = \sum_{\mbox{cycles: ijk...}} \alpha_R Tr R_{ijk...} + \alpha_S Tr S_{ijk...}
\label{cycles}
\ff

This is a first sketch at a gauge invariant dynamics, much needs to be done.

\section{Conclusions}

The aim of this paper was to provide a critical sketch of the foundational theory of neo-classical economics, and deduce from its strengths weaknesses how it may be improved upon by making use of ideas and techniques not available when the principles of that theory were formulated in the 1950s.  Since that foundational theory, the Arrow-Debreu model of economic equilibrium is precisely defined mathematically, it is straightforward to understand and critique.  It is not surprising, given the central place it has in modern ecnomic thinking, that it has considerable strengths.  But, as has been pointed out before by economists, it has a number of weaknesses.  We foused here on those issues having to do with the treatment of time and contingency.  These led to the conclusion that what is needed is a non-equilbrium and dynamical theory of economic markets by means of which we may deepen and extend our understanding of equilibrium in the Arrow-Debreu model.  By analogy with the central role of statistical mechanics in physics we suggest this might be called {\it statistical economics.}

We took a few baby steps towards a formulation of {\it statistical economics.}   While doing this we should keep in mind a great strength of the Arrow-Debreu model we discussed, which is its canonicalness and generality.  It is too easy to disregard this and strart playing with models arbitrarily picked from a large landscape of possible models.  One needs to work from principles which strongly constrain the forms of models we want to study.  

Nonetheless,  it seems likely progress can be made due to two opportunities not available to past economists.  The dramatic increases of computer power make it possible to study directly large agent based models.  The aim here is not to realistically model the details of real markets but to capture in the specification of relatively simple agents the key parameters and observables needed to understand the behavior of markets away from equilibrium.  Indeed many economists have been using agent based models in their research.  

We also followed Malaney and Weinstein in pointing out the key role that gauge invariance must play in formulations of economic models out of equilibrium.  Indeed, there is already gauge invariance in the Arrow-Debreu model, so this is not a new feature, but it appears that it should be extended to model the non-equilibrium market dynamics.  We argued that, as in physics, the principle of gauge invariance may strongly constrain the form of models of out of equilibrium dynamics of markets.  If this is correct then its exploration may lead to a non-equilibirum theory of markets as canonical and powerful as the Arrow-Debreu model of equilibrium.

\section*{ACKNOWLEDGEMENTS}

This work is the result of collaborations and conversations with a number of people. Conversations with Per Bak sparked my interest in economics and the example of his work remains an inspiration.   I am grateful to Stuart Kauffman and Mike Brown for invitations to collaborate, as well as to Jim Herriot, Zoe-Vonna Palmrose and Bruce Sawhill.  The  ideas described here evolved as a response to work of and with that group. Eric Weinstein has been extremely helpful at every stage of this work, that which is based on his joint work with Malaney, and more generally.  I am also very grateful to members of the  PI complexity/economics group for many discussions and insights,  especially Sundance Bilson-Thompson, Laurent Freidel, Philip Goyal, Sabine Hossenfelder, Kirsten Robinson, Simone Severini and Samuel Vazquez.  Several of them provided useful feedback on these notes, as did Robin  Hanson and Kelley John Rose.  Finally I learned important ideas from very helpful conversations and correspondence with Roumen Borissov, Jeffrey Epstein,  Richard Freeman, Thomas Homer-Dixon, Jaron Lanier, Roberto Mangabeira Unger, John Whalley, Steve Weinstein and Shunming Zhang. 

Research at Perimeter Institute for Theoretical Physics is supported in
part by the Government of Canada through NSERC and by the Province of
Ontario through MRI.

\end{document}